# Super-localization of excitons in carbon nanotubes at cryogenic temperature.

C. Raynaud,[1] T. Claude,[1] A. Borel,[1] M.-R. Amara,[1] A. Graf,[2]
J. Zaumseil,[2] J.S. Lauret,[3] Y. Chassagneux,[1] and C. Voisin[1]

[1]*Laboratoire de physique, École normale supérieure, PSL,*
*Université de Paris, Sorbonne Université, 75005 Paris, France*
[2]*Institute for Physical Chemistry, Heidelberg University, 69120 Heidelberg, Germany*
[3]*Laboratoire Aimé Cotton, École Normale Supérieure de Paris Saclay, Université Paris Saclay, 91400 Orsay, France*

At cryogenic temperature and at the single emitter level, the optical properties of single-wall carbon nanotubes depart drastically from that of a one-dimensional (1D) object. In fact, the (usually unintentional) localization of excitons in local potential wells leads to nearly 0D behaviors such as photon antibunching, spectral diffusion, inhomogeneous broadening, etc. Here, we present an hyperspectral imaging of this exciton self-localization effect at the single nanotube level using a super-resolved optical microscopy approach. We report on the statistical distribution of the traps localization, depth and width. We use a quasi-resonant photoluminescence excitation approach to probe the confined quantum states. Numerical simulations of the quantum states and exciton diffusion show that the excitonic states are deeply modified by the interface disorder inducing a remarkable discretization of the excitonic absorption spectrum and a quenching of the free 1D exciton absorption.

The optical properties of single-wall carbon nanotubes (CNTs) have triggered steady research efforts for more than a decade in the broader context of one-dimensional (1D) nanosystems [1] for applications in advanced photonics including sensing, bio-labeling, or recently single-photon generation for quantum cryptography [2–4]. Although the light emission properties of CNTs are now well understood at room temperature in the framework of one-dimensional excitons with high binding energy ($E_b \simeq 300$ meV, Bohr radius $R_b \simeq 1$ nm), this picture breaks down at cryogenic temperature. In fact, several studies have reported intriguing signatures such as massive luminescence line fragmentation at low temperature [5, 6], strongly asymmetrical photoluminescence (PL) lineshapes [7–9] or photon antibunching [10] that all point to localized excitons. This systematic spontaneous localization of excitons at low temperature is usually attributed to local inhomogeneities of the interface with the matrix or substrate or to unintentional adsorbates [6, 11] and is a fascinating consequence of the spill-out of the excitons wave function probing their local environment.

This ubiquitous behavior has very deep consequences in using CNTs for quantum-light generation because the change of paradigm from a 1D to 0D emitter yields drastic changes in the photon statistics, in the spectral purity, in the quantum yield and finally in the working wavelength of the source [4]. These consequences of unintentional exciton localization can be either beneficial or detrimental to applications. Nevertheless, the understanding of this spontaneous trapping mechanism and its implications remain elusive in the literature. Here, we make use of an innovative hyper-spectral imaging technique in the super-localization regime combined with quasi-resonant excitation spectroscopy to bring out a comprehensive picture of the excitonic states confinement and their coupling along the nanotube axis. Combined with numerical simu-lations of the confined quantum states in a random energy landscape and of the exciton diffusion dynamics, we were able to reproduce all of the experimental observations and to obtain a quantitative assessment of the potential energy disorder along the tube axis, including a coherence length of 20 nm and a standard deviation of 20 meV.

CoMoCat nanotubes were dispersed by shear force mixing in toluene and wrapped with the conjugated polymer PFO-BPy to prevent bundling effects and enhance the luminescence quantum yield [12]. This soft separation technique yields a relatively broad length distribution typically between 0.5 and 3 µm. The sample consists of individual (6,5) nanotubes ($d_t \simeq 0.76$ nm) randomly dispersed by spin coating on a fused quartz substrate coated with a 3 nm thick $Al_2O_3$ layer in order to protect the nanotubes from charge impurities at the interface [2]. The nanotube density is on the order of $0.03\,\mu m^{-2}$. The sample is placed on the cold finger of an optical cryostat and cooled down to 10 K.

A cw Ti:sapphire laser beam at 1.59 eV (non-resonant excitation) is focused on the sample (with an excitation density kept below $10\,kW/cm^2$) and the PL is collected through the same objective in a confocal configuration. The PL is separated from the excitation beam by a dichroic beam splitter and sent in a CCD spectrometer. PL maps are collected by coarse scanning the laser spot at the surface of the sample (Figure 1.a). Once a nanotube is located, a fine scan of a 1.5x1.5µm² area is carried out (Figure 1.b). The PL spectrum is recorded for each point of the map. A typical spectrum of a single nanotube (as inferred from the comparison with Raman or AFM maps, see SI) is shown in Figure 1.c). It displays a series of sharp lines in strong contrast with the PL of shorter nanotubes (200 nm or less) where single line emission is often observed [13]. For each line of the spectrum, a series of spectrally resolved maps is obtained by using narrow spectral integration windows (to avoid cross-talk between the lines) as indicated by the color code in figure 1.c. Strikingly, the spots obtained for each line are shifted with respect to each other (Figure 1.d-g).

The super-resolution technique is based on the assumption that the exciton localization length is much smaller than the diffraction limit. It was initially developed in the context of single molecule spectroscopy in dilute samples [14]. The position of the single emitter can be obtained with very high accuracy by fitting the spot to the point spread function of the microscope. The ultimate precision on the position of the spot center depends only on the signal to noise ratio. In principle, recording the emission intensity for two pixels in each direction only is enough to localize the emitter with an arbitrary precision [15]. In the case of carbon nanotubes however, the noise is known to be super-Poissonian (with a variance growing faster than the count rate N) due to blinking phenomena [16]. Thus, it is not possible to reach an arbitrary precision by increasing the integration time nor the excitation power. The optimum integration conditions are reached for a count rate of about 10 times the read noise of the CCD camera (see SI). In addition, for such super-Poissonian noise, the precision becomes better for a larger number of points in the PL map. Nevertheless, to avoid long-term thermal or mechanical drift of the setup, we used only 25 points in our measurements. Using independent methods based either on the analysis of the noise statistics or on the reproducibility of the maps, we obtained an estimate of the localization precision of about 15 nm (see SI).

The typical separation between the spectrally resolved spot centers (on the order of 100 nm) turned out to be much larger that the precision on their localization (∼15nm), therefore making it possible to visualize the spontaneous localization of excitons along the nanotube axis (figure 1.h), in contrast to previous studies that were carried out either with diffraction limited precision or with near-field technique that are prone to perturb the exciton localization. In order to check whether the spatial distribution of the localization sites is consistent with the topography of a single CNTs, we measured the local orientation of the nanotube axis by means of polarization measurements. In fact, it is well established that the luminescence of carbon nanotubes is highly polarized with a direction parallel to the tube axis [5]. We measured the polarization diagram for each of the lines of the PL spectrum. The polarization contrast is of the order of 30 which yields a precision of a few degrees in the local orientation of the nanotube. The polarization diagram of each line is reported on the map of the exciton localization positions in figure 1.h, which results in a consistent picture of a single CNTs with an average "vertical" orientation in this particular case. We reproduced those steps on a large number of individual carbon nanotubes and obtained statistical distributions of several physical quantities related to the trapping mechanism.

The distribution of the distance between nearest neighbors has a mean of 110 nm and a standard deviation of 100 nm (see SI). This reflects the random distribution of the excitonic traps along the nanotube axis together with an exclusion length that could physically correspond to the spatial extent of the traps (catch length). The long lasting tail can be attributed to overlooked PL peaks (overlapping or weakly emitting spectral features), which artificially increases the distance between two sites. Taking into account that the ends of a CNT are known to act as non-radiative defects [17], this mean distance between excitonic traps is consistent with the observation that nanotubes shorter than 200 nm are prone to single line emission at low temperature, as reported consistently in the literature [7, 8].

We are now interested in assessing the width and depth of these traps. The former can be obtained indirectly from the lineshape of the phonon side-bands that are observable when the spectral separation between the lines is large enough (typically above 10 meV, see SI). This is more likely to occur for shorter nanotubes (whereas for longer nanotubes, the spectral window will be congested with almost degenerate spectral lines). In fact, as reported in several studies, the typical PL profile at low temperature consists in a sharp zero-phonon line (the ZPL) together with asymmetrical phonon wings [8, 9, 18]. The width of these phonon wings is inversely proportional to the extension of the exciton center-of-mass envelope function, which is directly related to the trap width. We have extracted the exciton localization length from the fitting of the phonon wings for several lines measured on several individual nanotubes (see SI). We find consistently a localization width between 4 and 6 nm. This length is much larger than the exciton Bohr radius (on the order of 1 nm [19]), but much smaller than the diffraction limit and confirms our hypothesis of a point like emitter in the analysis of the maps.

The depth of the traps can be assessed indirectly from the energy of the PL lines. In fact, since all the nanotubes investigated in this work belong to the same chiral species and thus have the same band gap, the energy spread between the lines reflects the distribution of the traps depths. Since the trap width is typically much larger than the exciton Bohr radius, the confinement energy is small compared to the exciton binding energy. Therefore, the energy difference between the free and trapped exciton PL lines is in first approximation a good estimate of the trap depth. Nevertheless, the assessment of the free exciton energy is not straightforward. Figure 2 shows the segmentation of the PL spectrum of a CNT upon cooling from room temperature to 10 K. At low temperature, the lines spread over 70 meV but there is no line that can be undoubtedly attributed to the free exciton. This can be understood by the fact that its intensity becomes negligible if the trapping of exciton is very efficient. The zero energy level can be estimated from the center of the room temperature PL line by adding a shift corresponding to the estimated bandgap change upon cooling. For (6,5) nanotubes the latter is about +10 meV (blue shift) [20]. In total, the trap depth is given by the difference between the room temperature energy plus 10 meV and the low-temperature PL energy. A statistical study was carried out on several tens of nanotubes showing that the typical trap depth is on the order of 10 to 35 meV [6]. This means that the thermal detrapping probability is very low at 10K, but becomes likely at room temperature. This is consistent with previous studies showing excitons diffusing over several hundreds of nm at room temperature [21].

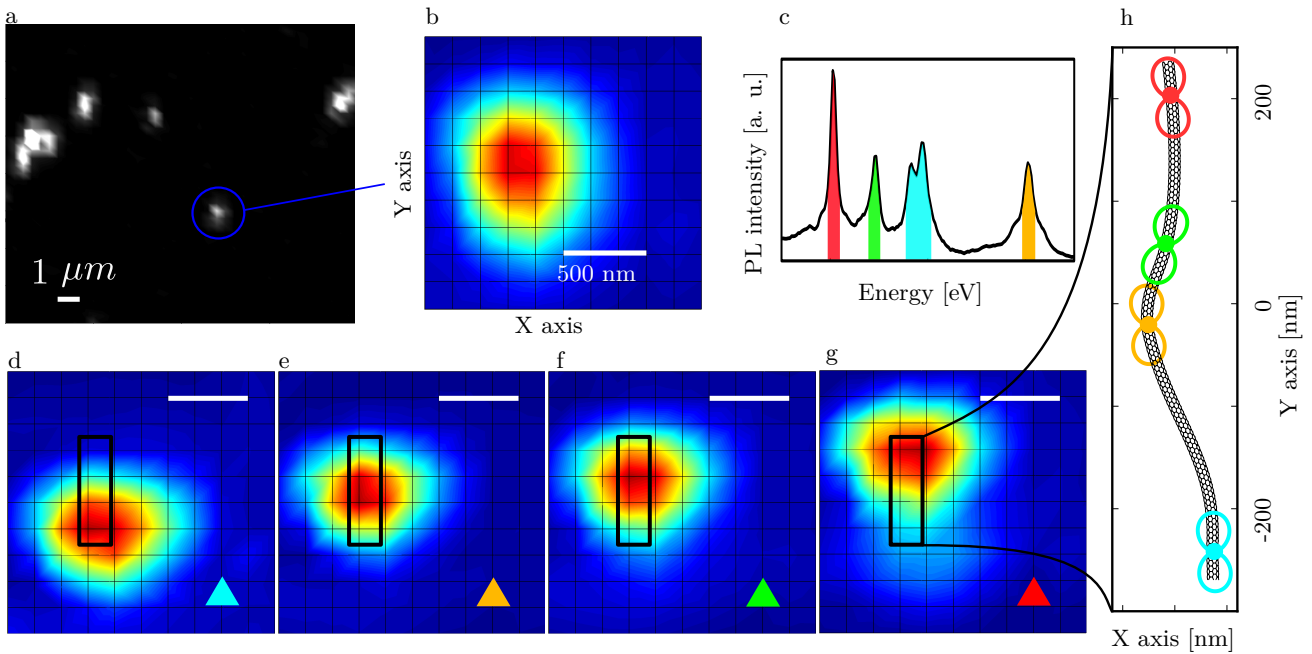

Figure 1. Superlocalization principle: (a) Coarse scan of the sample with the excitation beam. (b) Fine scan of the 1.7x1.7μm² area highlighted in (a). (c) PL spectrum of the nanotube observed in (b) for an excitation laser energy of 1.59 eV. (d-g) Excitation map of the PL of the same nanotube for a spectral filtering of the luminescence according to the color code of (c). (h) Locations of the centroid of each excitation spot (d-g) in the xy plan, together with the polarization diagrams measured for each PL line, indicating the local orientation of the nanotube. A schematic representation of the nanotube is added as a guide to the eye (the diameter of the tube is not at scale).

Nevertheless, we notice that a few traps seem to have a depth significantly larger that 25 meV, which is consistent with recent reports showing a partial antibunching of the PL emitted by a single nanotube at room temperature, meaning that part of this emission still arises from localized states even at room temperature [22]. This also means that the picture of the excitonic emission at room temperature is rather that of an inhomogeneous line with part of the emission due to freely diffusing excitons and the other part due to localized states.

Figure 2 shows another intriguing feature at low temperature : the presence of PL lines at higher energy than the mean energy at room temperature. Although this case is quite marginal (few lines of moderate intensity), it is an interesting signature of the energy landscape in which the exciton evolves. In particular, this observation questions the usual picture of a free exciton encountering a few localized traps, but is rather in favour of a random energy landscape showing sections at higher energies than the mean. These sections are not totally depopulated by diffusion due to energy barriers that segment the nanotube into quite independent parts at low temperature.

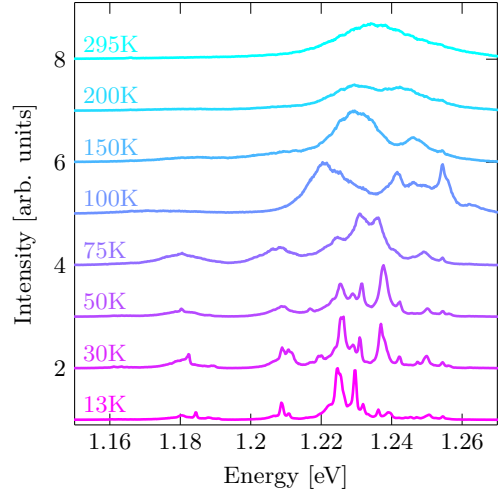

Figure 2. PL spectrum of a CNT recorded at several temperatures, showing the strong segmentation of the PL spectrum over a 50 meV window. Excitation energy 1.59 eV.

Finally, we briefly discuss possible physical origins for the excitonic traps. Importantly, we note that the traps are usually completely reconfigured once the sample is heated up to room temperature and cooled down again (see SI). This means that the energy needed to change the local environment of the nanotube is lower or on the order of 25 meV. We speculate that this is related to tiny changes in the conformation of the polymer wrapping the nanotubes (PFO-BPy). In fact, the exciton energy is very sensitive to changes in the dielectric permittivity in the very close vicinity of the tube due to the spill-out of the exciton wave-function. Ando *et al.* have shown that this exciton energy change can reach up to 90 meV considering the case of a surrounding with $\epsilon \sim 2.5$ whose interface with the nanotube would show a roughness of the order of a few angstroms only [23, 24].

Generally, the luminescence in semi-conductors arises from the lowest-energy defect states, whereas the excitation spectrum mainly reflects the large intrinsic density of states of the bulk that feeds the defect states through non radiative relaxation. In the case of carbon nanotubes, the question is whether excitation occurs through the "bulk" (delocalized) excitonic states or whether excitation rather involves absorption on localized excited states within the traps. We used a tunable cw Ti:sapphire laser [1.21 eV to 1.65 eV], which covers both the $S_{11}$ transition band and the main phonon side-band of the (6,5) chirality. This allows us to study the quasi-resonant photoluminescence excitation spectrum of single CNTs at low temperature (Figure 3). Experimentally, the laser energy is tuned by steps of 1.5 meV and a PL spectrum is taken at each step. For each PL line, the intensity is monitored as a function of the excitation energy and normalized to the incoming laser intensity. This brings valuable information about the excited states coupled to each trapped excitonic state. At relatively high excitation energy (about 100 meV above the PL energy), the excitation spectra of all the PL lines look rather similar (Inset of Figure 3 and SI). It mainly consists in a relatively broad resonance about 200 meV above the PL energy, known as the $X_2$ phonon side-band involving the zone-edge dark $KK'$ excitons [25, 26]. This exciton having a large center-of-mass momentum, it is by nature delocalized and it is thus consistently coupled to all trapped states. Note that the $X_2$ resonance overlaps well with the side bands observed at room temperature in the ensemble absorption spectrum in line with the interpretation in terms of delocalized states.

Using sharp low-pass filters, we were able to perform quasi-resonant excitation of the PL lines. For each line, we observed a series of sharp and strong excitation resonances in an energy window corresponding to the envelope of the room temperature absorption line ($S_{11}$ transition, see Figure 3). In contrast to the phonon side band, these resonances appear at different energies for the different PL lines of the same nanotube. This is in strong contrast with the simple model derived from usual semi-conductors where a few traps along the tube axis would be fed by a common free exciton absorption line : this scenario would indeed give rise to a common excitation line for all the PL features (see SI). Our observations rather mean that each localized excitonic states is coupled to a specific set of localized excited states. Plotting the excitation spectrum as a function of the detuning (Inset of Figure 3) does not show any constant energy shift between the emitting and absorbing states that would be the signature of a specific phonon assisted mechanism. We typically observe between 2 and 4 excitation resonances for each PL line. However, in the narrow overlap window of the PL and PLE spectra, we do not see any steady match between the PL lines and the excitation resonances. This is consistent with the picture of uncoupled potential traps along the nanotube axis. In this picture, the excitation resonances could correspond to the excited states in the trap. In fact, considering the simplified model of an exciton of mass $m = 0.2m_0$ in a truncated parabolic trap (local minimum of potential) of depth $V_t = 20 - 50$ meV and width $a = 3 - 10$ nm, we find an interlevel spacing of $\hbar\omega_0 = \frac{\hbar}{a}\sqrt{\frac{V_t}{m}}$ on the order of 15-30 meV, which corresponds to about 2-3 confined states in the trap, in line with the experimental observations. An interesting outcome of this study is that it is possible to excite specifically one localized state of the nanotube using an appropriate excitation wavelength (figure 3.a where the PL spectrum becomes almost single line when using an excitation laser at 1.245 eV).

This study brings out a consistent picture of exciton localization and light-matter interaction in carbon nanotubes at cryogenic temperatures. In fact, we were able to reproduce most of the experimental observations using a simple numerical model based on thermally assisted diffusion of the exciton along the nanotube in a random energetic landscape. We simulated the PL and PLE spectra of a 1 μm long nanotube, where the exciton energy landscape is randomly modulated with a Gaussian distribution of energies with a standard deviation of about 20 meV and a spatial coherence length of 16 nm. In the approximation where the electron-hole Coulomb interaction ($\sim 300$ meV) remains much larger than the confining potential, we solved numerically the Schrödinger equation for the exciton center of mass in the effective mass approximation (using $m^* = 0.2m_0$) and obtained the ground and excited confined states energies and envelope functions. The relative strength $f_n$ of the light-matter coupling for each of these states is given by the square of the integral of the envelope function [27, 28]. We first checked that the distribution of distances between neighbor energy minima in such an energy landscape is consistent with the experimental distribution (of mean 110 nm), taking into account that a certain fraction of the lines are overlooked in real experimental conditions (see SI). Note that the actual mean distance between traps in this simulated

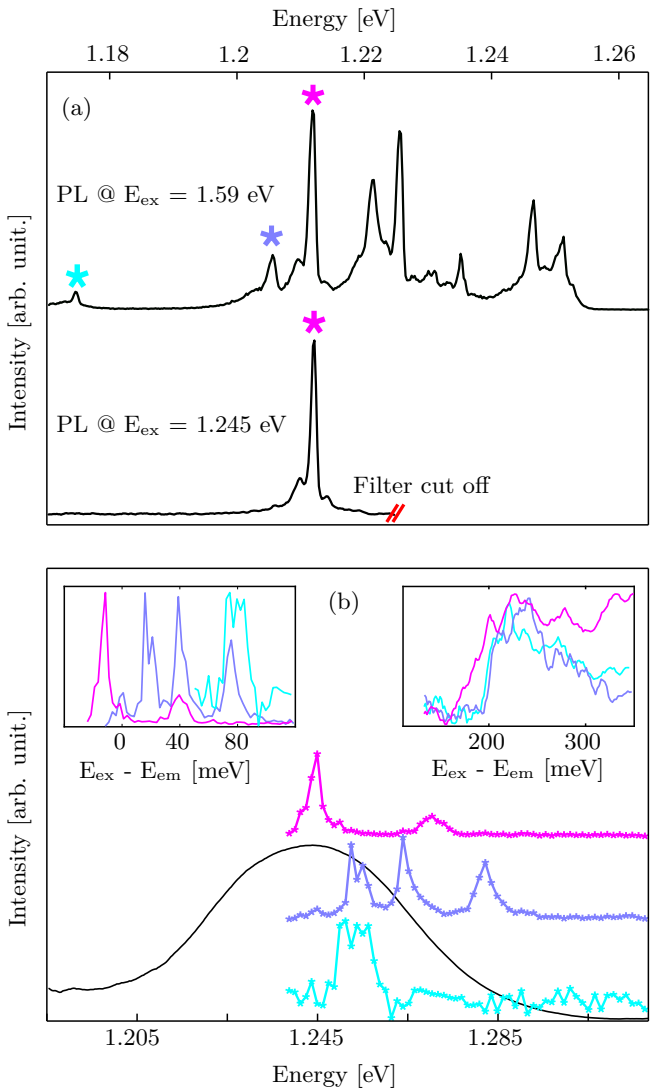

Figure 3. (a) PL spectrum obtained for non-resonant excitation at 1.59 eV (filter cut-off at 1.26 eV, vertically shifted black line) and for a selective excitation at 1.245 eV (filter cut-off at 1.22 eV). (b) Quasi-resonant PLE spectra of the individual PL lines of the same single nanotube at 13 K. The color correspondence between the emission lines and their respective PLE spectrum is indicated by the stars in panel (a). The black solid line is the ensemble absorption spectrum. (left inset) Same PLE spectra plotted as a function of the detuning between the laser and the PL line. (right inset) Higher energy PLE spectra showing the $X_2$ phonon side band excitation line common to all the PL lines.

potential is 70 nm.

We first considered the case of a non resonant excitation (as achieved for instance using the $X_2$ phonon side-band) where excitons are initially randomly distributed along the nanotube. Following the approach of Georgi *et al.* [29], we simulated the diffusion and trapping of the exciton using a hopping probability proportional to the Boltzmann factor of the energy difference between neighbour sites.

For each temporal step, a recombination probability is considered. The PL spectrum is obtained by summing up Lorentzian lines centered at the energy of the state occupied by the exciton when it recombined (see SI). The temperature broadening is accounted for by taking an effective Lorentzian width of $25T/300$ (meV), where $T$ is the temperature. This expression interpolates the observed low and room temperature linewidths. Figure 4.a shows the evolution of the simulated PL spectrum as a function of temperature. At low temperature, we obtain a set of narrow lines corresponding to the emission of the localized ground states in the random potential. Although tiny lines are observable above the zero (mean) energy (in agreement with experimental data), the emission is globally red-shifted as expected when the exciton population is frozen in the deepest traps. As the temperature increases, the lines broaden and the distribution of PL lines becomes more balanced across the whole energy span of the potential landscape. All these predictions are in qualitative agreement with the experimental data (Figure 2). Note that the excited states represent a marginal contribution to the PL spectra due to their reduced oscillator strength compared to the ground state and reduced occupation factor.

In a random potential, the effective diffusion length depends strongly on the temperature as the excitons become more prone to trapping at low temperature. We predict that the low temperature diffusion length is determined by the mean inter-trap distance and that it increases steadily with temperature up to a plateau given by the intrinsic diffusion coefficient and exciton lifetime at high temperature (see SI). Globally, the model explains why red detuned lines have an increasing intensity at lower temperature, which is consistent with the picture of excitons being gradually trapped in a more efficient way. Nevertheless, the intensity distribution of the lines at low temperature is not directly depicted by a Boltzmann distribution because of the uniform population feeding mechanisms along the tube axis and reduced effective diffusion length at lower temperature.

Next, we simulated the low temperature PLE spectra by considering the absorption of the excited states. The absorption strength is proportional to $f_n$. The energy width of the localized excited states is fixed to 2 meV, as deduced from the experimental data (Figure 3). At low temperature, each created exciton is supposed to relax to the nearest trap and to give rise to a PL line corresponding to the ground state of this trap. Figure 4.c shows the simulated PLE spectra for the 5 mains lines of the PL spectrum calculated in Figure 4.b, in a spectral window corresponding to our experimental conditions. We observe on average 2 to 4 excitation resonances that in general do not match the PL lines present in the same spectral range (dashed line in Figure 4.c). This exemplifies that ground and excited states play a fundamentally different role in the optical properties of CNTs at low temperature.

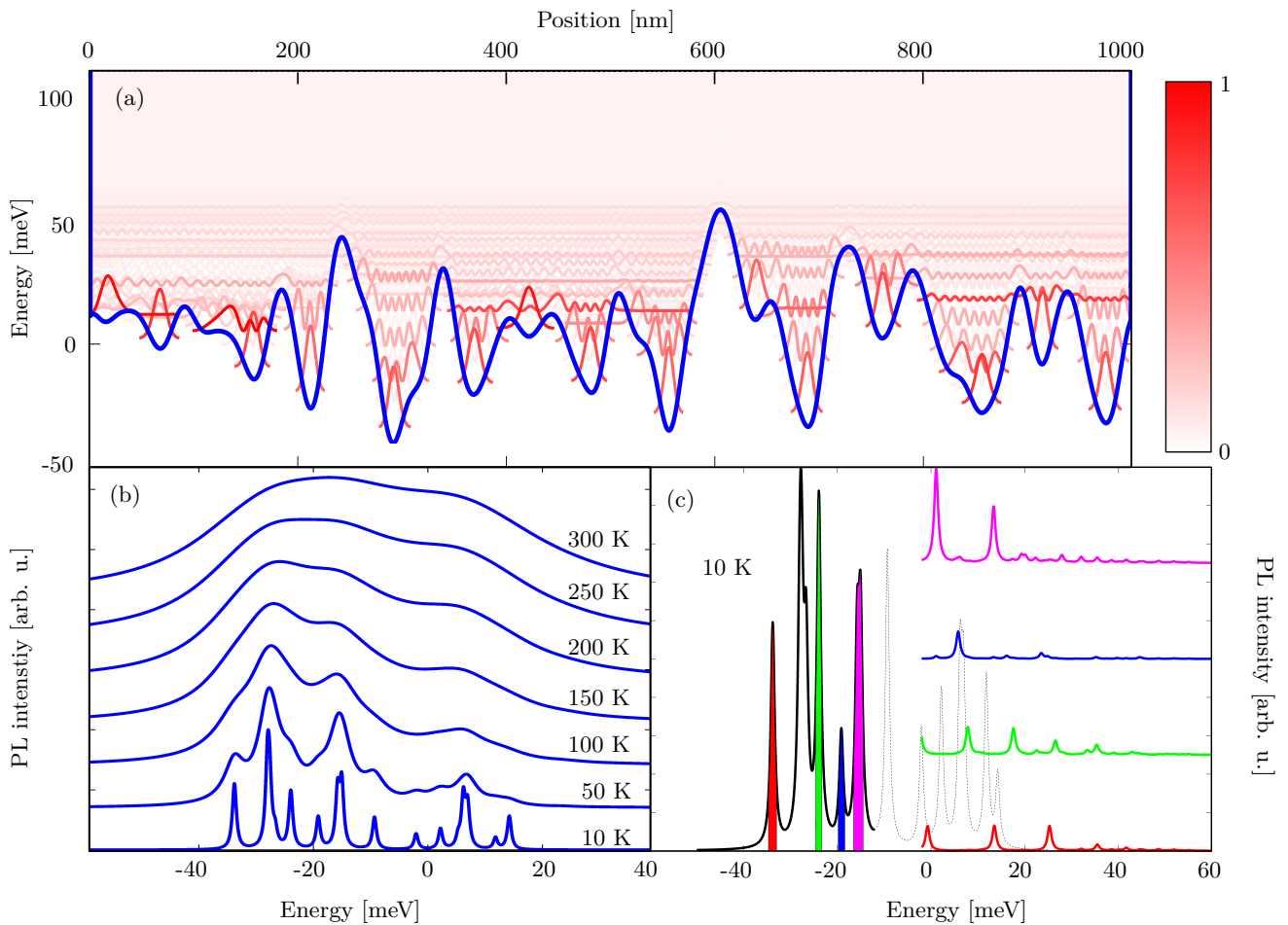

Figure 4. (a) Numerical simulation of the energy landscape for the exciton using a correlation length of 16 nm and a Gaussian distribution of the energy with a mean of 0 and a standard deviation of 20 meV. The localized quantum states are plotted with a vertical shift according to their eigenenergy and with a color code showing the strength of their coupling to the electromagnetic field (see SI). (b) Simulation of the resulting PL spectra for several temperatures, assuming a random initial population distribution along the nanotube (non-resonant pumping) and a diffusion coefficient of 900 nm² ps⁻¹. The exciton diffusion is simulated using a numerical resolution of the Fokker-Planck equation with a drift term reflecting the Boltzmann probability of the scattering events along the potential. The PL intensity reflects the integrated probability for the exciton recombination to occur in a given quantum state. (c) Simulation of the low temperature PLE spectra. For a given excitation energy, the exciton generation on a quantum state is proportional to its Lorentzian resonant factor (assuming a uniform width of 2 meV for all excited states and to the coupling strength of each state (see SI). The exciton is thus supposed to relax down to the nearest ground state cascading in the local energy gradient.

This also shows that the limited diffusion length of the exciton at low temperature makes the nanotube equivalent to a collection of almost independent localized emitters (quantum dots). Note that the localization of excitons in this disordered system essentially stems from dissipative processes (fast relaxation of excited states) rather than from coherent effects like Anderson localization effects.

In conclusion, we have shown that the roughness of the interface between the nanotube and the surrounding matrix yields a drastic randomization of the energy landscape for the exciton due to the probing of the local dielectric permittivity through the spill-out of the exciton wavefunction. At low temperature, this inhomogeneous potential energy leads to a systematic localization of the exciton and a segmentation of the PL line. Exploiting the spectroscopic signatures of the localization (including the nearly resonant excitation regime) and a superlocalization technique, we could estimate the mean distance between traps in this random potential to be on the order of 70 nm whereas the energy spread is on the order of 20 meV (Gaussian distribution). Notably, this model yields a straightforward explanation to the consistent observation that most short nanotubes (below 300 nm) display a single atomic-like PL spectrum. In addition, a key signature of this random interface potential is the disappearance of the $S_{11}$ free exciton absorption line at low temperature together with the rise of a series of sharp resonances corresponding to the excited states of the local potential wells. In total, from the viewpoint of optics, the nanotube behaves like an original emitter consisting in a collection a pseudo two-level systems that can be addressed -within the same diffraction limited spot- either individually through the localized excited states or collectively through higher energy 1D states. This approach could be generalized to create densely packed and individually addressable single-photon sources within the same nanotube using the emerging defect engineering techniques.

---

# Supplementary Information to : Super-localization of excitons in carbon nanotubes at cryogenic temperature.

C. Raynaud,[1] T. Claude,[1] A. Borel,[1] M.-R. Amara,[1] A. Graf,[2]
J. Zaumseil,[2] J.S. Lauret,[3] Y. Chassagneux,[1] and C. Voisin[1]

[1]Laboratoire de physique, École normale supérieure, PSL,
Université de Paris, Sorbonne Université, 75005 Paris, France
[2]Institute for Physical Chemistry, Heidelberg University, 69120 Heidelberg, Germany
[3]Laboratoire Aimé Cotton, École Normale Supérieure de Paris Saclay, Université Paris Saclay, 91400 Orsay, France

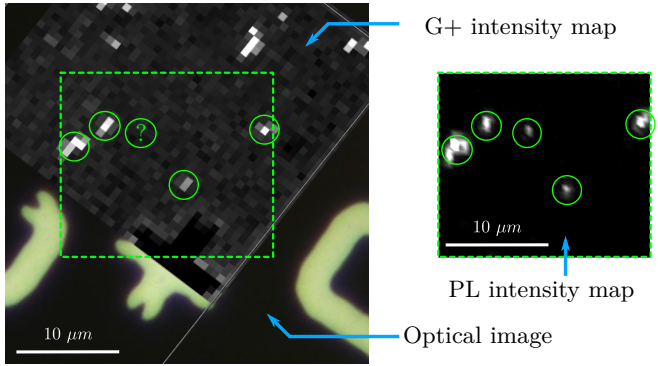

Figure S-1. Map of the G+ Raman mode recorded with a He-Ne laser (632.8 nm) superimposed to an optical image of the sample. PL map of the area in the dashed rectangle for a detection window of [1.2-1.25eV].

## I. RAMAN MAPS

The almost mono-chiral (6,5) nanotubes are spin-coated on a quartz substrate marked with gold labels that allow for a reproducible scanning of the same sample area with different techniques including Raman scattering and PL. Since all nanotubes belong to the same chiral family (in agreement with a uniform RBM mode at $310 \, \text{cm}^{-1}$), they are all equally excited at 633 nm and the Raman map directly reflects the nanotube distribution. Fig. S-1 shows the matching between PL and Raman maps.

## II. TEMPERATURE INDUCED RECONFIGURATION OF THE TRAPS

A strong indication that the trapping of excitons along the nanotube axis is not due to deep structural defects but rather to soft inhomogeneities of the nanotube/matrix interface is given by the steady observation of a drastic change in the PL spectrum of a given nanotube when the sample is brought back to room temperature and cooled down again to 10 K (Figure S-2). This suggests a deep reconfiguration of the energy landscape along the nanotube. Thus, the exciton traps must have a typical binding energy lower than 25 meV, which is rather in favor of conformational changes in the polymer matrix at the interface with the nanotube.

## III. SUPER-POISSONIAN NOISE CONTRIBUTION

Figure S-3 shows the time evolution of the PL intensity of one spectrally filtered line of a CNT. The time bin is 1s. The histogram of the intensity distribution is shown on the right panel together with the simulated Poissonian intensity distribution with identical mean count rate (blue), showing the super-Poissonian contribution in the data. The total variance of the experimental distribution is fitted to $\sigma^2 = e^2 + N + \alpha N^2$, where $e$ is a background noise and $\alpha$ is the super-Poissonian coefficient. Values for $\alpha$ are tube dependent but are typically found between $10^{-2}$ and $10^{-1}$.

## IV. LOCALIZATION ACCURACY

The super-localization techniques rely on the fact that a point-like emitter can be located with an arbitrary precision if the (diffraction limited) point-spread function of the microscope can be measured with an arbitrarily high signal-to-noise ratio. In the case of carbon nanotubes however, the super-Poissonian statistics of the emission sets a limit to this signal to noise ratio even for long integration time or large excitation power. The super-Poissonian contribution to the intensity fluctuations are estimated from the time-trace of the luminescence intensity of a CNT spectral line as discussed above. From fits on numerical simulations of Gaussian spatial intensity profiles with such noise, it is then possible to estimate the standard deviation on the position of the center of the spot as a function of the mean count rate and for several super-Poissonian coefficients (Fig. S-4.a). It turns out that the precision on the position of the emitter decreases with increasing non-Poissonian coefficient. In addition, the accuracy levels off for count rates larger than a threshold value related to the dark-count and to the non-Poissonian coefficient. In total, in our experimental conditions, the typical accuracy is on the order of 15 nm.

Fig. S-4.b shows the influence of the number of pixels in PL maps on the accuracy on the emitter position. In

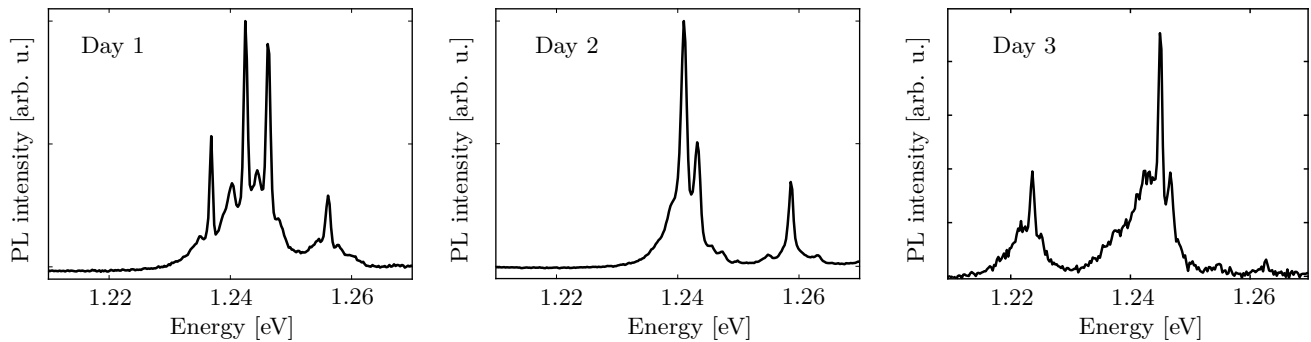

Figure S-2. Series of PL spectra of the same individual nanotube recorded at 10 K with the same experimental parameters after a cycle of heating up to room temperature and cooling down again to 10 K. The pressure in kept below $1 \times 10^{-3}$ mbar during the whole cycle.

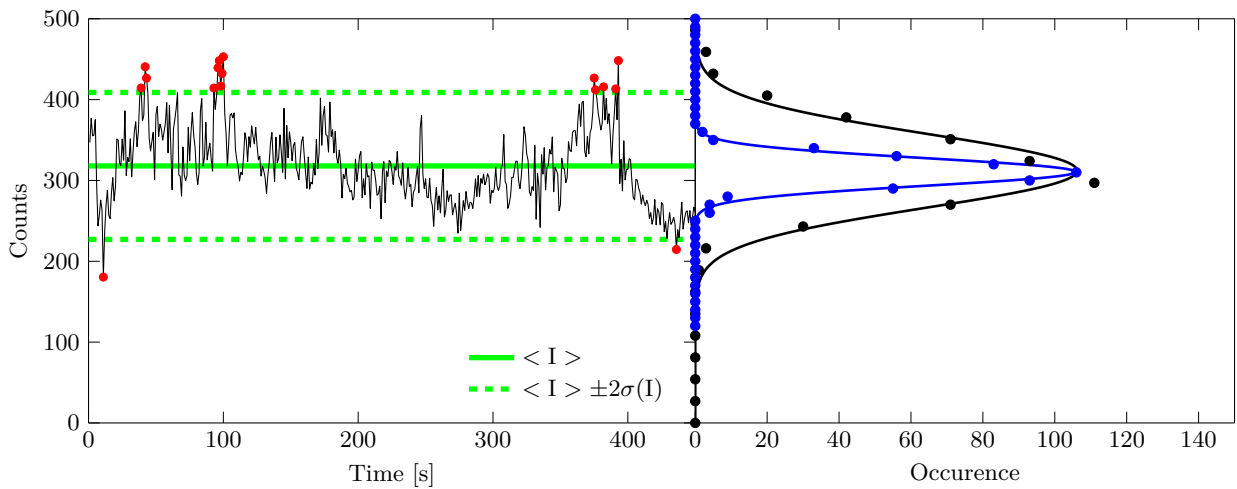

Figure S-3. (left) PL intensity of a spectral line of a CNT as a function of time (time bin 1s). The dashed lines show the $2\sigma$ limits of the count rate distribution. (Right) Histogram of the intensity together with Gaussian fit (black). Simulated Poissonian histogram for the same mean count together with a Gaussian fit (blue).

contrast to the Poissonian case, where the best accuracy is obtained whatever the number of pixels (N>2 in each direction), in the presence of a non-Poissonian component in the noise, the precision is greatly enhanced when a larger number of pixels is used in the PL map. As a trade-off between noise and long term thermal or mechanical drift, we used 5x5=25 pixels.

Finally, this estimate of the super-localization accuracy was benchmarked against reproducibility measurements. By taking several consecutive maps of the same luminescence spot, we confirmed that the spot center localization accuracy is below 20 nm.

## V. SPATIAL DISTRIBUTION OF THE TRAPS

In Figure S-5, we display additional maps of exciton trapping along the tube axis through super-localization measurements, together with the overall histogram of the distance between neighbouring traps. The mean of this distribution is 110 nm and the standard deviation is 100 nm. The inset shows the simulated trap distribution extracted from the random potential used to generate the spectra of Figure 4 of the main text. The distribution yields an average inter trap distance of 70 nm, shorter than the experimental one. However, assuming that a fraction of the order of 0.5 of the traps are overlooked in real experimental conditions (line overlap, low luminescence yield...), the apparent trap distribution becomes very similar to the experimental one.

## VI. TRAP WIDTH AND PHONON WINGS ANALYSIS

Based on the non-Ohmic exciton-phonon coupling model developed in [1, 2], we obtained the size of the center-of-mass exciton envelope function $\sigma$ from a fit of the PL profile including the phonon wings (Figure S-6). Basically, the exciton size is inversely proportional to

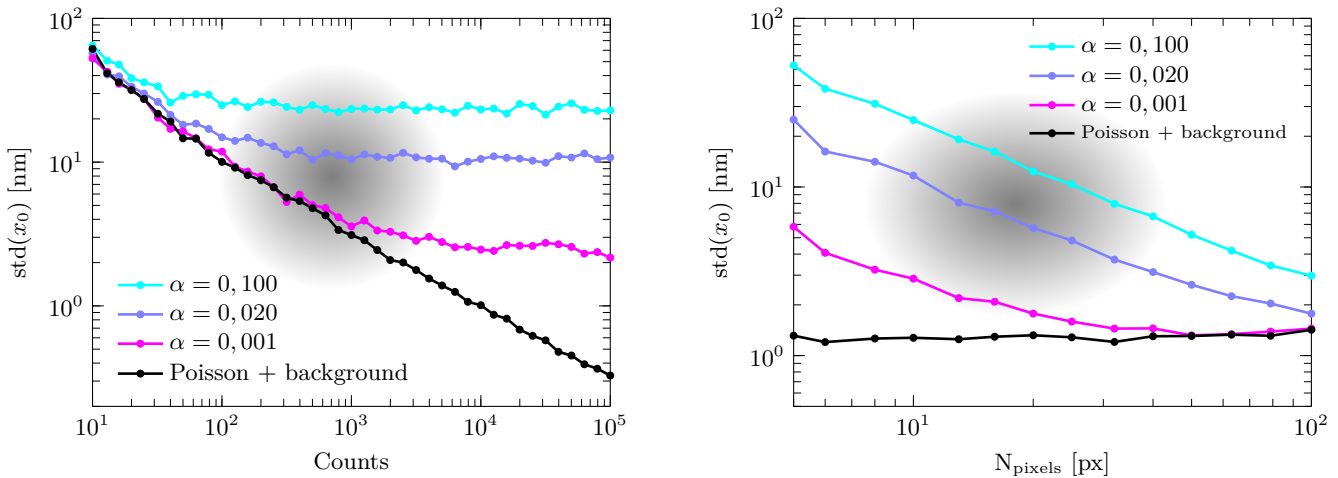

Figure S-4. (left) Standard deviation of the super localization technique as a function of the number of detected photons for several values of the super-Poissonian coefficient in the noise (the number of pixels is 11 on both the x an y directions). In contrast to the Poissonian case, super-Poissonian photon emission leads to a saturating precision above a certain count rate. (right) : influence of the number of pixels (in each x and y directions) on the super-localization resolution for several super-Poissonian contributions for a total count of $10^4$.

In both panels, the experimental conditions are depicted by the shaded area.

the spectral width of the phonon side-bands. This stems from the energy/momentum conservation in the phonon assisted exciton recombination and yields a cut-off Stokes shift for the phonon side-band of $\hbar v_s/\sigma$ where $v_s$ is the acoustic phonon velocity. We consistently get values of $\sigma$ between 4 and 6 nm, in agreement with previous reports. Note that this value is consistent with the modeling presented in the main text and in the section below. In fact, when fitting by a Gaussian the square modulus of the ground state wave-functions in each trap, we obtain an average exciton size of 5.5 nm.

## VII. EXCITATION SPECTROSCOPY

We provide additional examples of PLE spectra for quasi-resonant (figure S-8) and phonon side-band (figure S-7) excitation.

## VIII. MODELING

In a first attempt to model the exciton creation, diffusion and recombination in the nanotubes, we used a model where a few localized traps are distributed along the nanotube axis. The energy baseline corresponds to the free exciton. This model gives a sizable contribution of the free exciton to the PL spectra and a strong contribution of the free exciton to all PLE spectra, in contrast to the experimental data (Figure S-9).

Next, we used a totally random potential landscape $V(x)$, defined by its correlation function $C(x') = \overline{V(x)V(x+x')}$. We chose a Gaussian correlation function

$C(x') = (\Delta E)^2 \ \exp\left(-\frac{x'^2}{2l_{coh}^2}\right)$. To obtain such a potential, we add Gaussian white noise in the Fourier domain of $V(x)$:

$$V(x) = \Delta E \sqrt{\frac{l_{coh}}{2\sqrt{2\pi}}} \int dk \ e^{ikx} e^{-k^2 l_{coh}^2/4} \big(r_1(k) + i \ r_2(k)\big),$$

where $r_1$ and $r_2$ are Gaussian random variables with unitary standard deviation, and null mean value. In order to have a real potential $V(x)$, $r_1$ is even with $k$ and $r_2$ is odd. Correlations reads : $\overline{r_1(k)r_1(k')} = \delta(k-k') + \delta(k+k')$, $\overline{r_2(k)r_2(k')} = \delta(k - k') - \delta(k + k')$ and $\overline{r_1(k)r_2(k')} = 0$. The values of $\Delta E$ and $l_{coh}$ are adjusted in order to have a FWHM of 40 meV in the distribution of the local minima of energy, and a mean distance between neighbour wells of 70 nm. We found that $\Delta E = 20$ meV and $l_{coh} = 16$ nm match these conditions. Numerically, the distribution was obtained following the method given in [3].

The exciton binding energy being typically 10 times larger than the confinement energy, the electron-hole interaction is treated first and the role of the random potential is accounted for in the effective mass approximation for the exciton center of mass. We solved numerically the 1D Schödinger equation associated to this potential, using a mass of $0.2 \times m_0$ associated to the exciton center of mass. The numerical resolution of the Shrödinger equation is done by finding the eigenvector (wavefunctions) and the eigenvalues (energies) of the following $n$ x $n$ matrix (where $n$ is the number of points along the tube axis, corresponding to a discretization step $\delta x = 0.1$ nm ; in

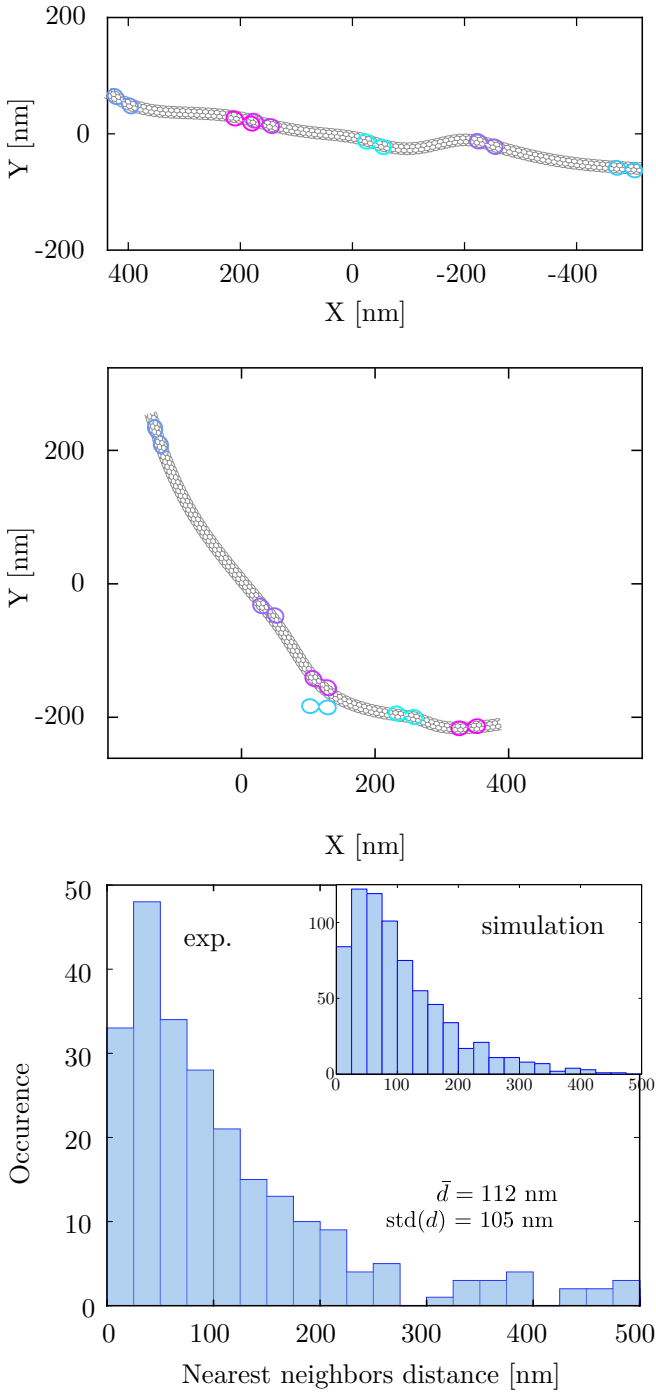

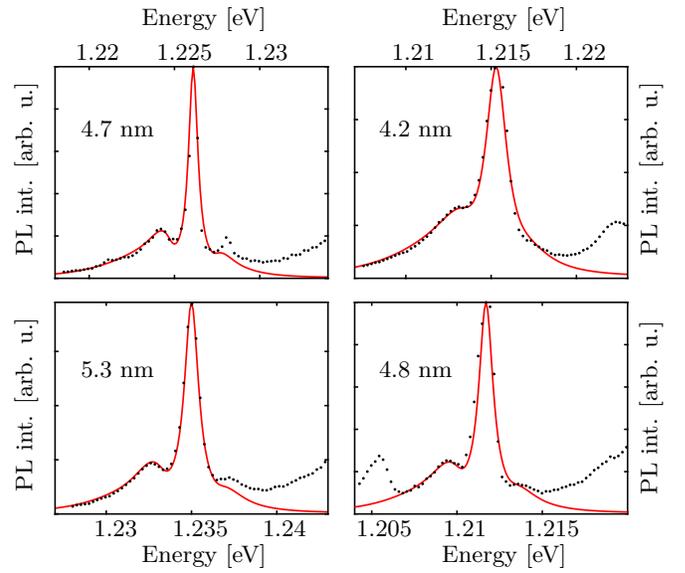

Figure S-6. Selection of PL lines of a CNT at 10 K where the phonon side bands are well resolved. Fits of these PL profiles to the non-Ohmic model as developed in [1, 2] (red lines). The width $\sigma$ of the exciton center-of-mass envelope deduced from these fits is indicated in black (Gaussian model). Note that the lesser agreement with the blue wing is due to partial overlap with neighbour lines.

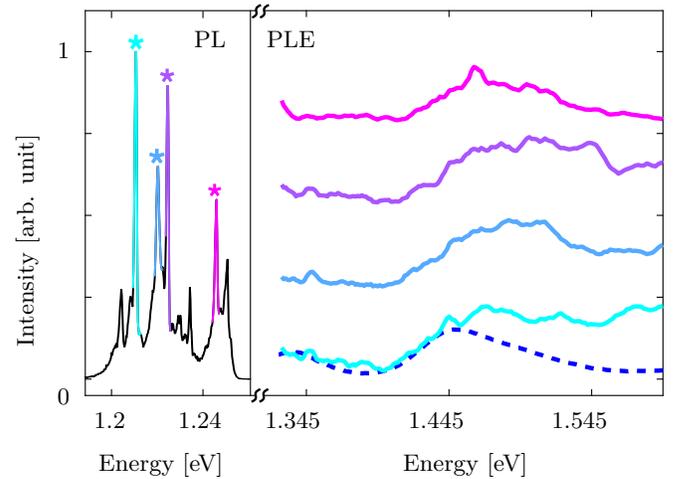

Figure S-7. Photoluminescence excitation spectrum of each emission line of a single CNT at 15 K, together with the PL spectrum (black) obtained for non resonant excitation (1.59 eV) and with the absorption spectrum (measured on an ensemble, dashed blue). This nanotube is the same as in Figure 3 of the main text but additional lines are investigated.

Figure S-5. (a-b) Additional examples of the measurement of exciton trapping along the nanotube axis by means of superlocalization technique together with polarization diagrams indicating the local orientation of the nanotube. The representation of the nanotube is only a guide for the eyes. (c) Histogram of the spatial distance between nearest neighbors. The bin corresponds to the experimental resolution. Inset : simulated trap distribution for the random potential used in Figure 5 (main text) and assuming that about 50% of the trap states are not detected in experimental conditions.

the simulation we used $n = 10000$):

$$\mathbf{H} = -\frac{\hbar^2}{2m^*}\frac{1}{\delta x^2}\mathbf{A} + \mathbf{V},\qquad(2)$$

where $\mathbf{A}$ is the square matrix with $-2$ value on the diagonal and $+1$ on the first superdiagonal and subdiagonal.

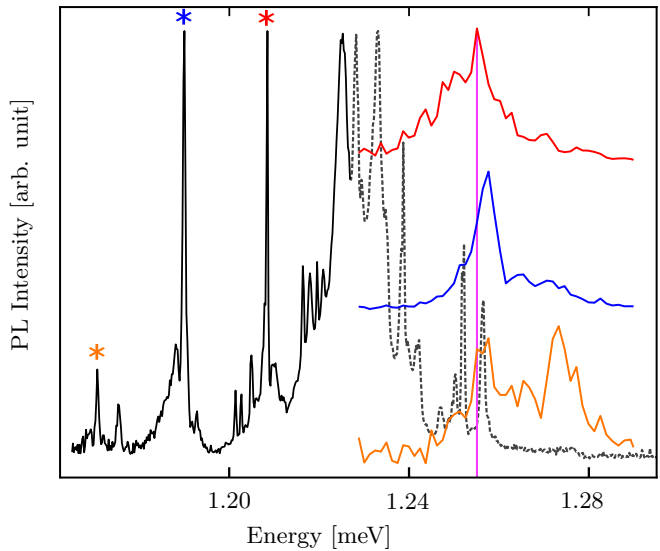

Figure S-8. PLE spectra of another nanotube showing localized states resonances. (see main text)

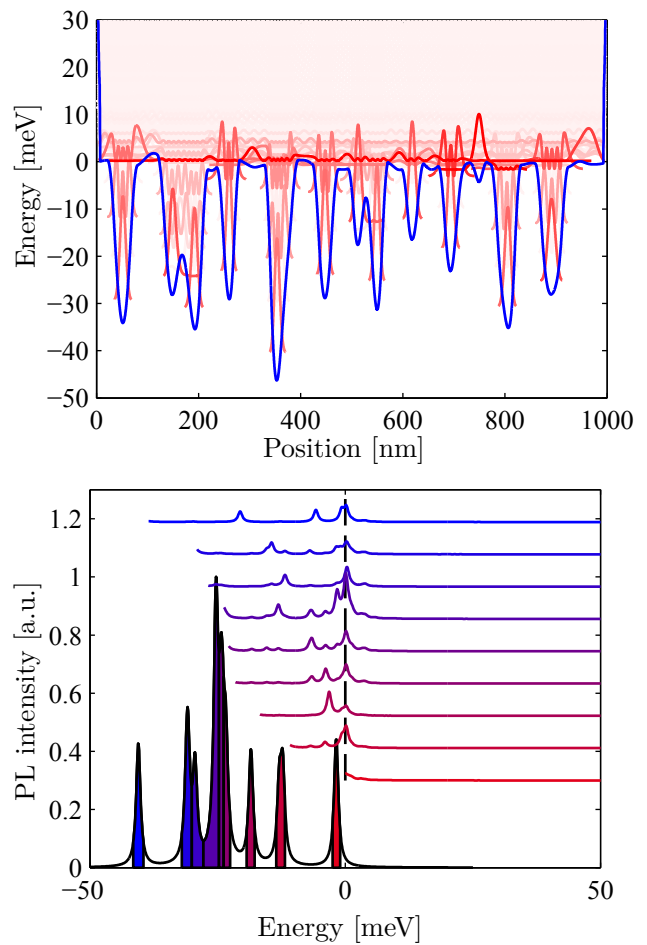

Figure S-9. Localized traps along the energy baseline of the free exciton in a single carbon nanotube. Corresponding simulated PL and PLE spectra showing a steady free exciton line in the PL spectrum and a strong contribution of the free exciton to all PLE spectra.

This matrix corresponds to discrete version of the operator $\frac{\partial^2}{\partial x^2}$. The matrix $\mathbf{V}$ is diagonal with values corresponding to $V(x)$. The eigen quantum levels are shown in Figure 4 of the main text. The wave function is displayed only in places where it takes significant values.

The color code shows the strength of the coupling to light. According to the textbook separation of relative and center-of-mass variables, it can be shown that the oscillator strength $f_i$ of the dipolar interaction with an incoming electromagnetic field is proportional to the square of the integral of the envelope wavefunction over the nanotube length: $f_i \propto \left| \int \psi_i(x) dx \right|^2$. Therefore, excited states which have a more pronounced oscillatory component are much less coupled to light than the ground states (for each trap). In addition their weaker average population at low temperature (Boltzmann factor) yields on average a negligible contribution of the excited states to the PL spectra.

In contrast, excited states play a key role in the quasi-resonant PLE spectra and explain the sharp and specific lines. In the low-temperature PLE simulations, the excitons generated on each excited states (at a rate proportional to the oscillator strength) are supposed to relax down to the lower ground state cascading in the local energy gradient. This relaxation is supposed to be much faster than the PL lifetime.

When raising the temperature the exciton are not completely trapped in local energy minima any longer and diffusion along the tube axis has to be taken into account. However, the typical inter-level spacing for confined states is on the order of 10 meV (Figure 4). Thus, it is not possible to evaluate the diffusion of the exciton using the continuous external potential $V(x)$ because the energy levels accessible to the exciton are discrete at the scale of

$k_B T$. The usual approach in this case is to use instead the Bohm quantum potential [4]. In this framework, the potential $U(x)$ felt by quantum particle is given by the energy of the quantum state having the highest density of probability at each position $x$ along the nanotube axis (Figure S-10).

Next, the diffusion of the exciton along the nanotube is computed by solving numerically the diffusion equation [5]:

$$\frac{\partial N}{\partial t} = D \frac{\partial^2 N}{\partial x^2} - \frac{1}{\tau} N - \frac{\partial (A(x)N)}{\partial x}, \qquad (3)$$

where $N(x,t)$ is the exciton density, $D$ is the diffusion coefficient and $\tau$ is the exciton lifetime. A drift term $\partial_x(AN)$ is used to describe the non-uniform potential. In the limit of infinite lifetime, the steady state exciton density $N_s$ should follow a Boltzmann distribution, which leads to $A(x) = \frac{D}{N_s} \frac{\partial N_s}{\partial x}$. In the discrete description, $A$

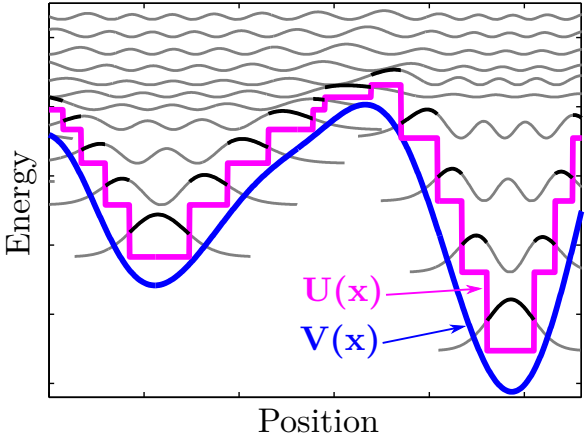

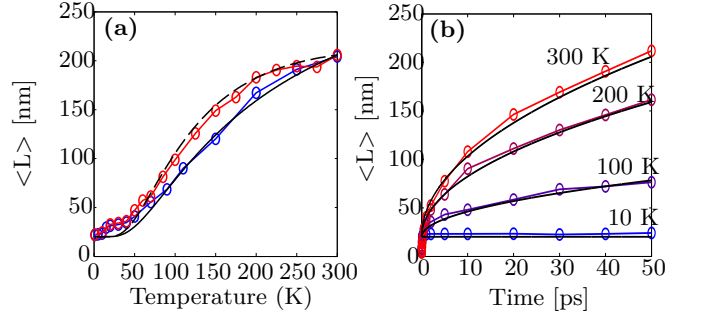

Figure S-11. (a) Simulation of the diffusion length as a function of temperature for a fixed lifetime of 50 ps (blue symbols), and for a temperature dependant lifetime (red symbols). The black lines are fits to eq. (9). (b) Evolution of the diffusion length as a function of time for temperatures of 10 K, 100, K 200 K and 300 K. The black lines are fits to eq. (9).

Figure S-10. Transformation of the potential for diffusion simulations. The wavefunctions (in black) are calculated using the original potential $V(x)$. For each position the wavefunction with the highest density of probability is drawn by a thicker line. Their corresponding eigen energy are used to generate the new potential $U(x)$.

reads:

$$A(x) = \frac{2D}{\Delta x} \frac{e^{-U(x-\Delta x/2)/kT} - e^{-U(x+\Delta x/2)/kT}}{e^{-U(x-\Delta x/2)/kT} + e^{-U(x+\Delta x/2)/kT}} \quad (4)$$

The exciton population is defined on each position $i\Delta x$, and the potential on $i\Delta x + \Delta x/2$. By using $\partial_{x^2}N = (N(x+\Delta x) + N(x-\Delta x) - 2N(x))/(\Delta x)^2$ and $\partial_x(AN) = (A(x+\Delta x)N(x+\Delta x) - A(x-\Delta x)N(x-\Delta x))/(2\Delta x)$, the diffusion equation reads :

$$\begin{aligned} N(x, t+\Delta t) = &-\frac{\Delta t}{\tau}N(x,t) + p_s N(x,t) \\ &+ p_l(x+\Delta x)N(x+\Delta x) \\ &+ p_r(x-\Delta x)N(x-\Delta x), \quad (5) \end{aligned}$$

where $p_s$, $p_l(x)$, and $p_r(x)$ are, respectively, the probability for the exciton of staying at the same position, of going one step left and of going one step right. These probabilities read :

$$p_s = 1 - 2\frac{D\Delta t}{\Delta x^2} \quad (6)$$

$$p_l(x) = \frac{D\Delta t}{\Delta x^2}(1 - B(x)) \quad (7)$$

$$p_r(x) = \frac{D\Delta t}{\Delta x^2}(1 + B(x)) \quad (8)$$

where $B(x) = \Delta x A(x)/(2D)$. As expected, the ratio $p_r/p_l = \exp(-\Delta U/kT)$ is simply the Boltzmann factor.

In the simulation we use $\Delta x = 0.3$ nm, and $dt = 0.25dx^2/(2D)$. We use a temperature dependant lifetime given by $\tau = 50/(0.8 * T/300 + 0.2)$ ps, where $T$ stands for the temperature which is a good approximation of the values reported by Berger et al. [6]. The value of the diffusion constant $D$ is tuned in order to have a diffusion length of 200 nm at room temperature [7].

In the simulations, excitons are launched one by one in a random initial position. The diffusion length is obtained by calculating the average distance travelled by the exciton in a lifetime. We found that a value of $D = 900\text{nm}^2\text{ps}^{-1}$ yields $< L > \approx 200$ nm at 300 K. We stress that the diffusion length is smaller than ($< L_{flat} > = \sqrt{2D\tau} = 300$ nm). This is a consequence of the rough energy profile that reduces the excursion length in the exciton random walk. By changing the temperature and the lifetime $\tau$ in the simulations, (Figure S-11), we find that the diffusion length is well approximated by :

$$< L > \approx \sqrt{2D_{eff}(T)\tau} + L_0/\sqrt{12} \quad (9)$$

The effective diffusion coefficient is given by an Arrhenius law $D_{eff} = D_0 \exp(-E/kT)$ [8], with an intrinsic value of the diffusion coefficient $D_0 = 1100$ nm$^2$/ps and $E = 30$ meV [9]. The offset term $L_0/\sqrt{12}$ is negligible at high temperature but becomes dominant at low temperature. This term corresponds to the average distance between traps $L_0 = 70$ nm.

The PL spectrum is calculated from the integral over time of the exciton population at each position. Each position along the nanotube is associated to a PL line at an energy given by the local value of the exciton energy and an intensity proportional to the oscillator strength $f_i$ calculated for the local quantum state associated to this energy. Numerically, we use equation (5) and let the population evolve for five lifetimes.

Finally, we validated the statistical parameters of the model by summing up the PL spectra obtained at 10 K for a set of 10000 realizations of the random potential. We compared this average spectrum to the experimental one obtained by summing up the spectra of 20 different nanotubes (Figure S-12.a). We observe that both the lineshape and the spectral width are in good qualitative agreement between simulations and experiments (number of peaks, asymmetry of the line, ...).

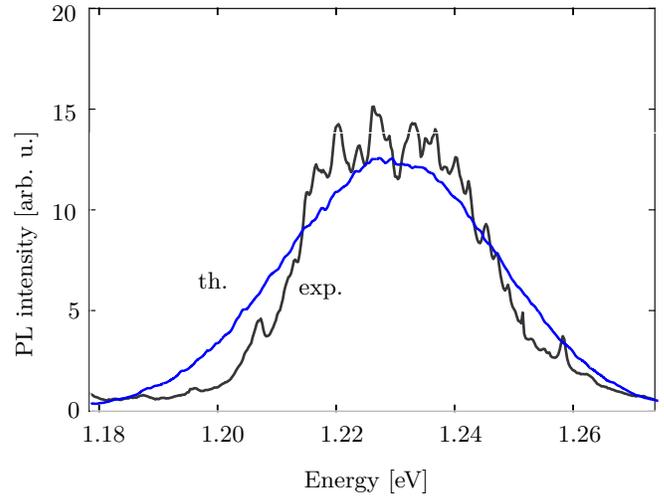

Figure S-12. Average low temperature spectra. The black curve corresponds to the sum of 20 spectra taken at 10K. The blue curve is the mean of 10000 simulated spectra at 0 K.